\documentclass[12pt,a4paper,oneside]{article}

\usepackage[top=1in,bottom=1in,left=.5in,right=.5in]{geometry}
\usepackage[margin=.5in,small]{caption}

\usepackage{authblk}
\usepackage{cite}
\usepackage{doi}
\usepackage{url}
\usepackage{amsmath,mathtools,siunitx,stackrel}
\usepackage{hyperref,cleveref}
\usepackage{epsfig}
\usepackage{graphics,graphicx}
\usepackage{amsfonts, amssymb, upgreek}
\usepackage{textcomp}
\usepackage[english]{babel}
\usepackage{blindtext}
\usepackage{color,xcolor}

\graphicspath{{img/}}

\newcommand{\tumble}{\mathtt{T}}
\newcommand{\run}{\mathtt{R}}
\newcommand{\xmotion}{\mathtt{X}}

\newcommand{\molar}{\mathrm{M}}
\newcommand{\Hill}{n_{_\mathrm{H}}}
\newcommand{\CCW}{_\mathrm{CCW}}
\newcommand{\CW}{_\mathrm{CW}}

\newcommand{\upd}{\mathop{}\!\mathrm{d}}

\newcommand{\ex}{\mathrm{e}}

\begin{document}

\bibliographystyle{plos2009}

\title{Stochastic model for the CheY-P molarity in the neighbourhood of \textit{E. coli} flagella motors}

\author[1]{G. Fier}
\author[1,2]{D. Hansmann\thanks{David.Hansmann@conicet.gov.ar}}
\author[1,2]{R. C. Buceta\thanks{rbuceta@mdp.edu.ar}}
\affil[1]{Instituto de Investigaciones F\'{\i}sicas de Mar del Plata, UNMdP and CONICET}
\affil[2]{Departamento de F\'{\i}sica, FCEyN, Universidad Nacional de Mar del Plata}
\affil[{ }]{Funes 3350, B7602AYL Mar del Plata, Argentina}

\maketitle

\abstract{\textit{Escherichia coli} serves as prototype for the study of peritrichous enteric bacteria that perform runs and tumbles alternately. Bacteria run forward as a result of the counterclockwise (CCW) rotation of their flagella bundle, which is located rearward, and perform tumbles when at least one of their flagella rotates clockwise (CW), moving away from the bundle. The flagella are hooked to molecular rotary motors of nanometric diameter able to make transitions between CCW and CW rotations that last up to one hundredth of a second. At the same time, flagella move or rotate the bacteria's body  microscopically during lapses that range between a tenth and ten seconds.  
We assume that the transitions between CCW and CW rotations occur solely by fluctuations of CheY-P molarity in the presence of two threshold values, and that a veto rule selects the run or tumble motions. We present Langevin equations for the CheY-P molarity in the vicinity of each molecular motor. This model allows to obtain the run- or tumble-time distribution as a linear combination of decreasing exponentials that is a function of the steady molarity of CheY-P in the neighbourhood of the molecular motor, which fits experimental data. In turn, if the internal signaling system is unstimulated, we show that the runtime distributions reach power-law behaviour, a characteristic of self-organized systems, in some time range and, afterwards, exponential cutoff. In addition, our model explains without any fitting parameters the ultrasensitivity of the flagella motors as a function of the steady state of CheY-P molarity. In addition, we show that the tumble bias for peritrichous bacterium has a similar sigmoid-shape to the CW bias, although shifted to lower concentrations when the flagella number increases. Thus, the increment in the flagella number allows lower operational values for each motor increasing amplification and robustness of the chemotatic signaling pathway.



\section{Introduction\label{sec:intro}}

Flagellated bacteria, when swimming in three-dimensional isotropic liquid media, execute motion modes, basically based on translations and/or turns, characteristic of each species. These movements may be limited by the interaction with their congeners or the surrounding material environment. The motion modes of flagellated bacteria are determined by the rotation of their flagella, which are hooked to a cation rotary motor. In multiflagellated species, the motors have the ability to be synchronized. The internal biochemical processes involved in the dynamics of the flagellar rotary motor at nanoscopic scale cause counterclockwise (CCW) and clockwise (CW) flagella rotations at microscopic scale. These internal processes are spatially located in the neighborhood of rotary motors and temporally trigger, quasi-instantaneously, the switching of flagella rotation direction.

Enteric peritrichous bacteria as \textit{Escherichia coli} have two alternating motion modes: persistent runs without setbacks (or recoils), and abrupt turns known as tumbles \cite{Berg1972, Macnab1972, Macnab1977b}. An \textit{E. coli} advances running forward when its flagella bundle (directed rearward) rotates CCW as a whole, and performs tumbles when as few as one flagellum rotates CW, detaching from the flagellar bundle  \cite{Turner2000}. In contrast, marine uniflagellated bacteria such as \textit{Vibrio alginolyticus} swim with cyclic motion modes established by persistent runs, first forward (with its flagellum behind) and then backwards (with its flagellum to the front), motion known as run reverse. This cycle is restarted by a sudden turn known as flick \cite{Taylor1974}. \textit{V. alginolyticus} advances running forward when its flagellum rotates CCW, and performs reverse runs when it flagellum rotates CW \cite{Homma1996}. Self-propelled microorganisms, and in particular flagellated bacteria, are systems that remain out of equilibrium, transiting between two or more metastable states, \textit{e.g. E. coli} transits between states of run and tumble and \textit{V. alginolyticus} transits between states of run and  run-reverse. Experimental and theoretical studies on motion of \textit{E. coli} have shown that tumbles cannot be considered instantaneous events \cite{Kafri2008,Saragosti2012,Fier2017,Fier2018}. Tumble-time distributions are a consequence of the internal signaling system and are essential in a complete description of the run-and-tumble motion. A conclusion drawn from fluorescence microscopy studies on \textit{E. coli} have concluded that the transition from run to tumble is the result of a change of direction from CCW to CW of at least one flagellar motor, which is visualized when one or more flagella move away from the flagella bundle \cite{Turner2000}. Recently, some very elaborated models have been introduced using a `veto' hypothesis, which assumes that the bacterium runs if all flagella rotate CCW and the bacterium tumbles if at least one flagellum rotates CW \cite{Vladimirov2010, Sneddon2012, Mears2014}. This veto hypothesis is based on high-definition observations using slow-motion techniques and optical tweezers \cite{Darnton2007}. However, until now it is unknown whether the veto hypothesis is sufficient to explain swimming behaviour, since the experimental measurements have not been reproduced theoretically from a model for the activity of flagellar motors at nanoscopic scale. Figure~\ref{fig:vote-hypothesis} shows a scheme of the veto hypothesis for a bacterium with three flagella running and tumbling.
\begin{figure}[ht]
\centering
\includegraphics[width=0.6\textwidth]{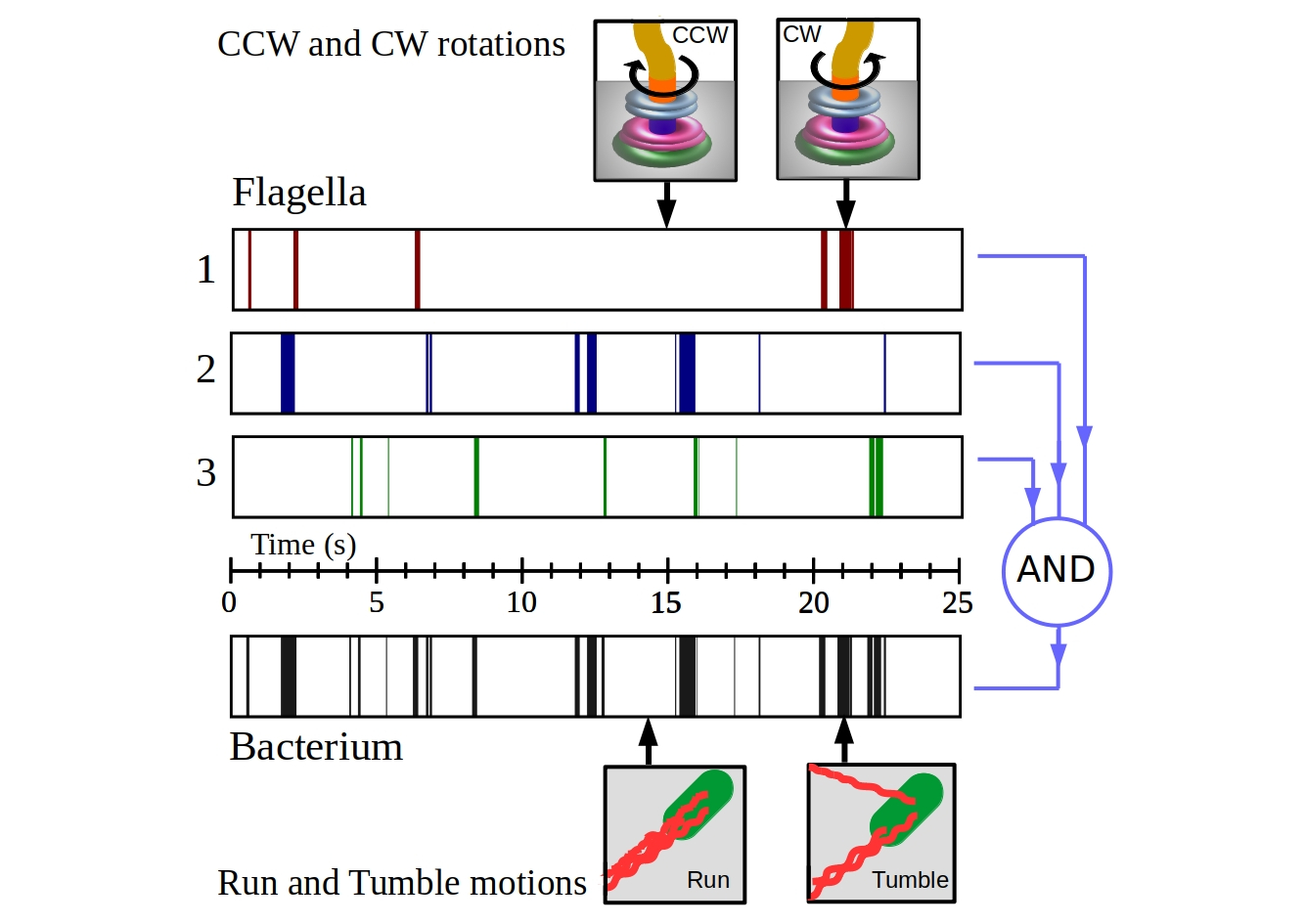}
\caption{The flow block diagram shows the veto hypothesis for a bacterium with 3 flagella from the data obtained from the numerical simulation of the internal signaling model introduced by us in Section~\ref{sec:two}. The 3 upper schematic timelines show the rotation intervals CW (coloured) and CCW (white). The veto hypothesis acts with the \texttt{AND} gate, which assumes that the bacterium runs if all its flagella rotate CCW. Otherwise, the bacterium tumbles. The lower schematic timeline shows the intervals of tumble (black) and run (white). Using the right-hand rule, the CCW and CW rotations have incoming (\texttt{i}) and outgoing (\texttt{o}) directions, respectively, seen from the flagellar motor.}
\label{fig:vote-hypothesis}
\end{figure}

The motion-modes behaviour of \textit{E.coli} is a consequence of the internal biochemical processes ocurring inside each bacteria. The signal-transduction system of histidine-aspartate phosphorelay consists, in a simplified fashion, of histidine-kinase transmembrane which activates response regulators proteins \mbox{CheY} and \mbox{CheB} on the cytoplasm \cite{Wadhams2004}. This signaling system responds to chemical stimuli and induces changes in bacterial behaviour, which are known as taxis. In turn, the response regulator \mbox{CheY} interacts with the subunits of the flagella producing transitions between runs or tumbles modes. However, the run-and-tumble motion in isotropic media is taxis free, although changes in the rotation direction of flagella occur without an external stimulus. Several results show that concentration of the CheY phosphorylated signaling protein (CheY-P), in the vicinity of the flagellar motor C-ring switch, determines the rotation direction of each flagellum \cite{Sourjik2002b}. In turn, the switch is composed of change proteins (FliN, FliG and FliM), with which CheY-P interacts to trigger the CCW-CW transition. It is known that the amount of FliM proteins involved in the transition CCW$\rightarrow\,$CW is not equal to the transition CW$\rightarrow\,$CCW \cite{Bren2001}. Additionally, chemical stimuli on the receptor cluster of the bacterium poles modify the levels of CheY-P on the cytoplasm and change the transitions between motion modes accordingly \cite{Sourjik2002b}. This phenomenon, known as chemotaxis, modifies only the frequency of the transitions between CCW and CW. Assuming that these transitions are caused by fluctuations of the CheY-P concentration between two threshold values in the neighborhood of the flagellar motors, we introduce a phenomenological stochastic equation for the CheY-P molarity in Section~\ref{sec:two}. This model allows to obtain the time distributions for run and tumble as a linear combination of decreasing exponentials, which fits experimental data. Also, the model shows that the runtime distributions follow a power-law behaviour when the internal signaling system is unstimulated.

An astonishing property of \textit{E.coli} is the gain of its chemotactic signaling system \cite{Segall1986}. Small variations in receptor occupancy lead to significant changes in the fraction of time in which the bacteria make runs or tumbles. This amplification has different origins. On the one hand, there is amplification due to receptor clustering in the cell poles. On the other hand, the ultrasensitive flagellar motors contribute to increase the amplification. The high cooperative response of the flagellar motors to variations in the molar concentration of CheY-P has been widely reported throughout the past decades \cite{Kuo1989,Cluzel2000,Sourjik2002b,Yuan2012,Yuan2013}. In Section~\ref{sec:two} we show that our model allows obtaining the CW bias as a function of steady CheY-P molarity in agreement with the experimental data. Besides, we show that the tumble bias is also a sigmoid function of steady molarity with the same slope as the CW bias, although it is displaced to its left. We conclude that another possible amplification factor is the flagella number. We show that, with the same ligand concentration, the tumble bias is approximately an order of magnitude greater than the CW bias. Additionally, a higher flagella number increases the robustness of the chemotactic response.

\section{Internal signaling system\label{sec:two}}

When \textit{E. coli} performs runs or tumbles in an isotropic media, its internal signaling system is unstimulated. Nevertheless, theoretical and experimental results show that the receptor-kinase-activity has a steady state in the abscence of stimuli \cite{Tu2008,Shimizu2010}. In consequence, the CheY-P molar concentration has a steady state under these conditions \cite{Sourjik2002a}. Under the assumption that the flagella motor has cilindrical geometry with a diameter of $\SI{45}{\nano\meter}$ and the bacterium has a cigar shape of a length of $1\,\mathrm{to}\,\SI{10}{\micro\meter}$ and a diameter of $\SI{1}{\micro\meter}$, the quotient between the area of a bacterium and the area of a motor is approximately in the interval $[2,20]\times 10^3$, which means that each \textit{E. coli} has enough surface for thousands of flagella but has only a few. In the light of this difference of scales between cell surface and the surface of the flagella motor, we model the concentration of CheY-P around the motor as a homogeneous stochastic process. We assume that transitions between CCW and CW occur by fluctuations in the concentration of CheY-P \cite{Morton1998} under the presence of two threshold values \cite{Bren2001}. Assuming that molarity evolves temporarily in a similar way around each flagellum motor, we introduce the following phenomenological Langevin equation (LE) for the molarity $c_j(t)$ of CheY-P in the surroundings of the $j$-th motor ($1\le j\le n$ integer with $n$ the flagella number):   
\begin{figure}[t]
\centering
\includegraphics[width=0.4\textwidth]{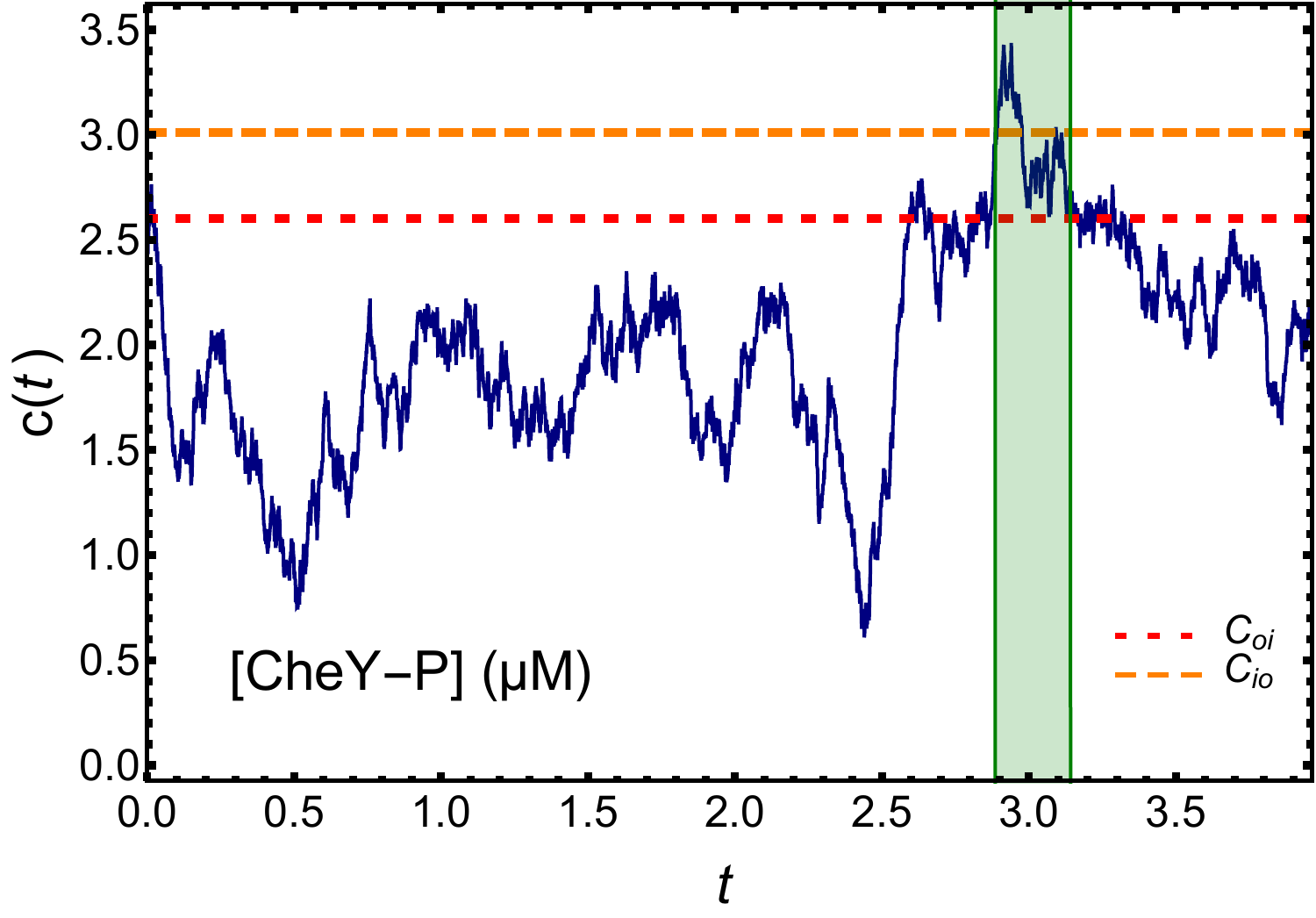}
\caption{The plot shows the molar concentration $c(t)$ of Chey-P as a function of time $t$ in the vicinity of one of the flagellar motors of \textit{E. coli}, obtained through the numerical integration of the Langevin equation~\eqref{eqn:c(t)}. Dotted lines show the molarity threshold-values $c_{\mathtt{io}}$ and $c_{\mathtt{oi}}$. If the molarity increases at $c(t) = c_{\mathtt{io}}$, there is a transition from CCW to CW, or else, if the molarity decreases at $c(t) = c_{\mathtt{oi}}$, there is a transition from CW to CCW. The coloured (colourless) regions correspond to the CW (CCW) regimes. The values of the integration parameters are $\beta =1$ and $\mu=\SI{2}{\micro\molar}$, while the chosen molarity threshold-values shown in the plot are $c_{\mathtt{oi}}=\SI{2.59}{\micro\molar}$ and $c_{\mathtt{io}}=\SI{3.01}{\micro\molar}$.}
\label{fig:chey-p}
\end{figure}
\begin{equation}  
\dot{c}_j=-(c_j-\mu)\,c_j^{\beta} + \zeta_j(t)\,,\label{eqn:c(t)}
\end{equation}
where $\dot{c}_j$ denotes time derivative of $c_j(t)$, $\mu$ is the steady molarity, and $\beta$ is a positive exponent to be determined. The noise $\zeta_j$ is Gaussian white with zero mean and correlations \mbox{$\langle\zeta_j(t)\zeta_k(t')\rangle=2\,Q_j\,\delta_{jk}\,\delta(t-t')$}, where $Q_j$ is the noise intensity and $j,k=1,\dots,n$. In order to numerically obtain a mean and variance close to that of the experimental results, it is sufficient to use a positive integer exponent $\beta$. The associated Fokker-Planck equation (FPE) to equation \eqref{eqn:c(t)} is
\begin{equation}
\frac{\partial}{\partial t}\,p(c_j,t|c_{j}^\prime, ṭ^\prime)=-\frac{\partial}{\partial c_j}[A(c_j)\,p(c_j,t|c_{j}^\prime,t^\prime)]+ Q_j\,\frac{\partial^2}{\partial c_j^{2}}\,p(c_j,t|c_{j}^\prime,t^\prime)\;,
\label{eqn:FPc}
\end{equation}  
with initial condition $p(c_j,t^\prime|c_j^\prime,t^\prime)=\delta(c_j-c_j^\prime)$, where $p(c_j,t|c_{j}^\prime,t^\prime)$ is the probability density function (PDF) of the molarity $c_j$ at time $t$ given the initial condition $c_j^\prime$ at time \mbox{$t=t^\prime$} and \mbox{$A(c_j)\doteq -(c_j-\mu)\,c_j^{\beta}$} is the drift of the process.

In the vicinity of the $j$-th motor, if the molarity increases at $c_j(t) = c_{\mathtt{io}}$, the transition from CCW to CW occurs; or else, if the molarity decreases at $c_j(t) = c_{\mathtt{oi}}$ (with $c_{\mathtt{oi}}<c_{\mathtt{io}}$), the transition from CW to CCW happens as shown in Figure~\ref{fig:chey-p}. The labels \texttt{i} (or \texttt{o}) denote that the flagellum performs CCW (or CW) rotation, that is, it moves with incoming (or outgoing) direction of rotation seen from the motor. During the CCW rotation, close to any motor, the probability [or survival probability function (SPF)] that the molarity has values $0< c(t)\le c_{\mathtt{io}}$ (with the initial condition $c(t^\prime)=c_{\mathtt{oi}}$) is 
\begin{figure}[ht!]
\centering
\includegraphics[scale=0.25]{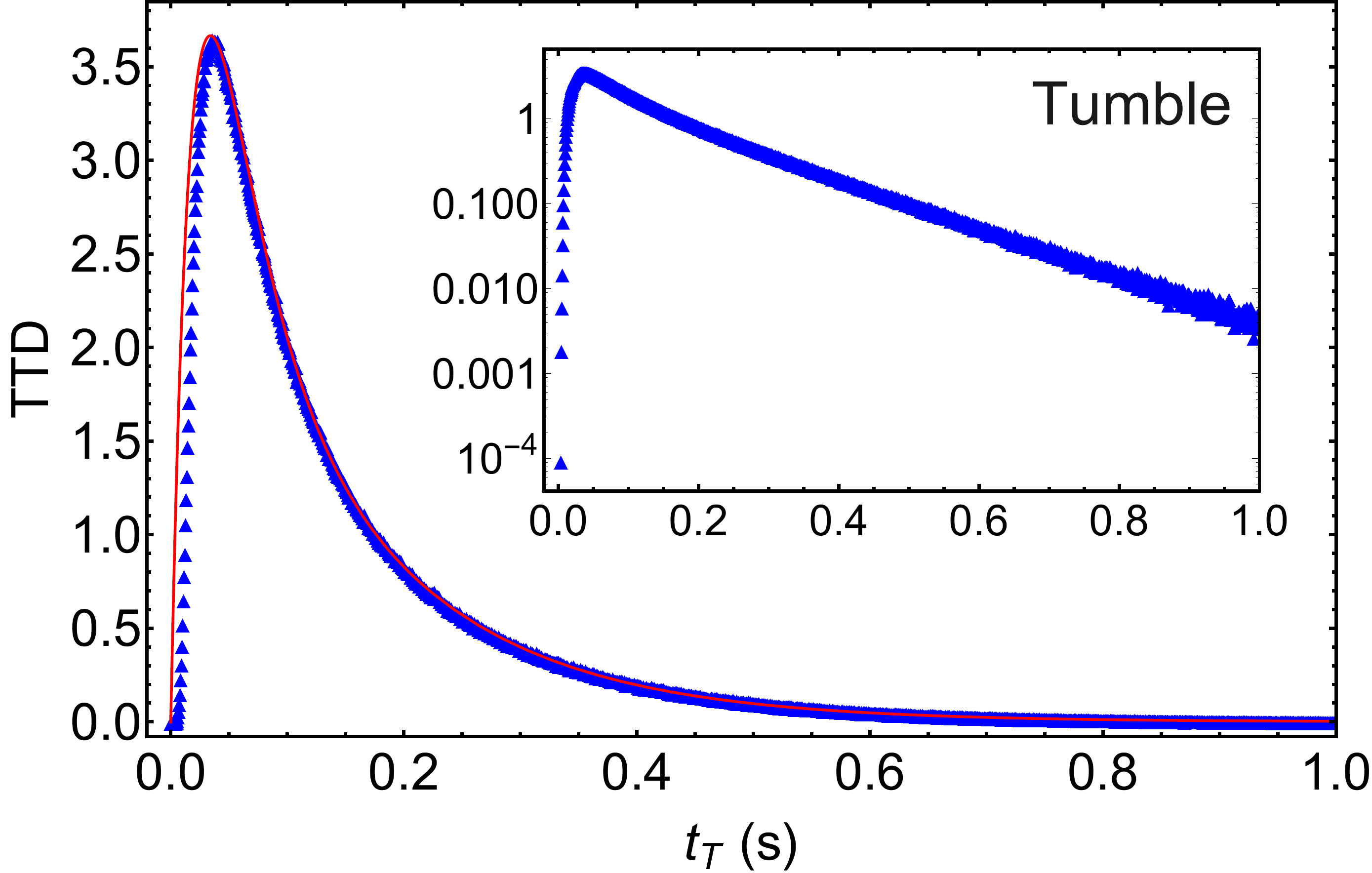}\hspace{1cm}
\includegraphics[scale=0.25]{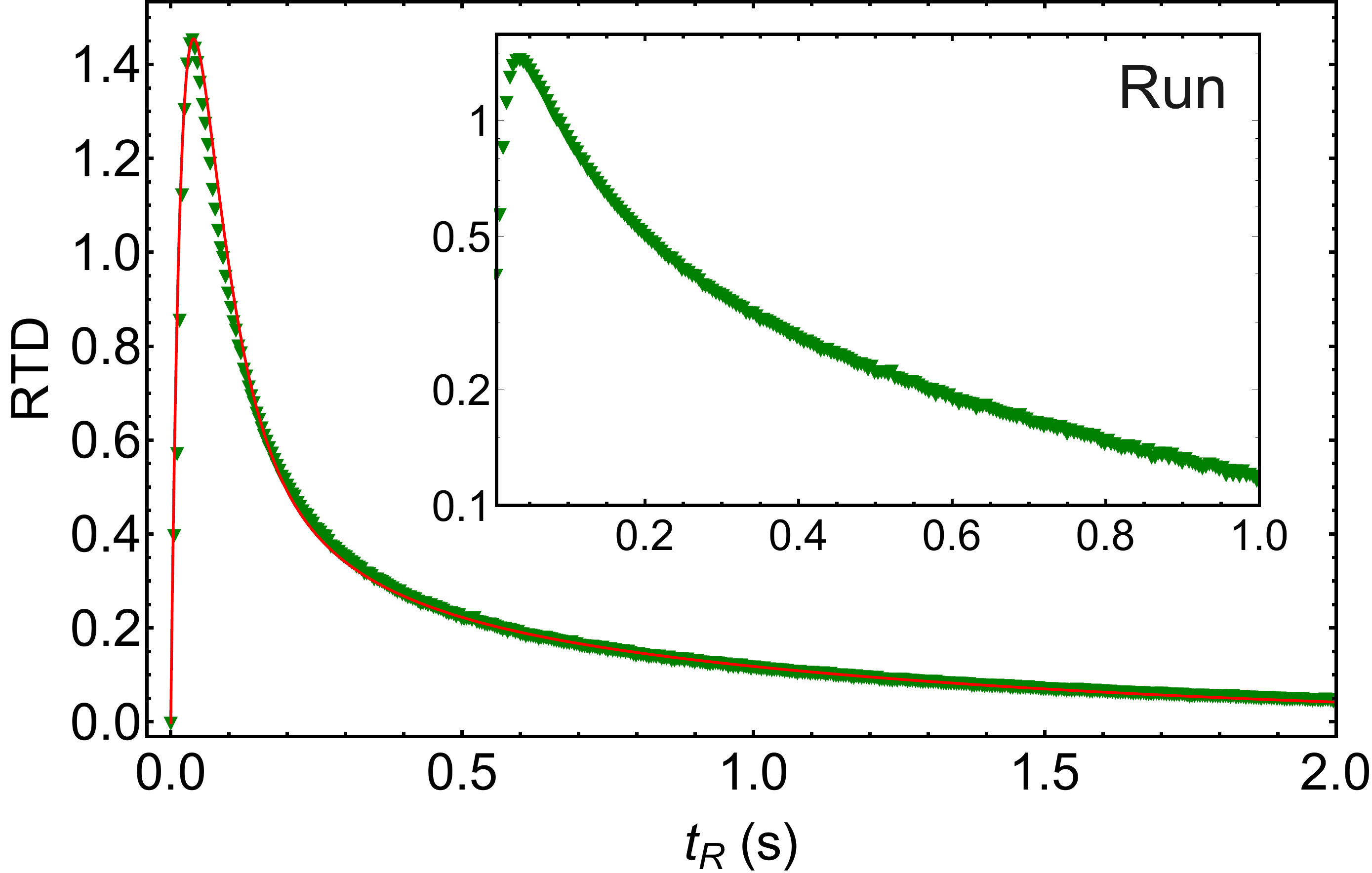}
\caption{Both plots show the tumble-time (TTD) and runtime (RTD) distributions as functions of tumble-time $t_\tumble$ and runtime $t_\run$ respectively, obtained from the internal signaling system model for a bacterium with three flagella ($n=3$). The plots with symbols are the result of numerical integration of Langevin equations~\eqref{eqn:c(t)} taking $\beta =1$, $\mu=\SI{2}{\micro\molar}$, and $Q_j=0.7$ ($j=1,2,3$)\,. The steady molarity value corresponds to an unstimulated internal signaling system. The chosen molarity threshold-values are $c_{\mathtt{oi}}=\SI{2.59}{\micro\molar}$ and $c_{\mathtt{io}}=\SI{3.01}{\micro\molar}$. Inside, we show the  semilog plots in order to see that the tumble-time distribution after the maximum is very well approximated by a single decreasing exponential function, behaviour that does not happen for the runtime distribution. In both plots, the maxima appear at times $\approx\SI{0.04}{\second}$ and the time distributions are equal to zero at initial times. The solid line corresponds to the data fit using a linear combination of exponential functions. The temporal mean values with their standard errors obtained from our simulations are $\SI{0.86}{}\pm\SI{1.21}{\second}$ for runs and $\SI{0.146}{}\pm\SI{0.136}{\second}$ for tumbles, which are close to the ones obtained for \textit{E. coli} \cite{Berg1972}.} 
\label{fig:TimeDistr}
\end{figure}
\begin{equation}
S_\mathtt{i}(\tau)=\int_0^{c_{\mathtt{io}}} \breve{p}(c,t|c_{\mathtt{oi}}, t^\prime)\,\upd c\;,
\label{eqn:RSPF}
\end{equation}
where $\tau=t-t^\prime$ and
\begin{equation}
\breve{p}(c,t|c^\prime,t^\prime)=p(c,t|c^\prime,t^\prime)-p_\mathrm{st}(c)\;,
\end{equation}
being $p_\mathrm{st}(c)$ the stationary PDF. In addition, during the CW rotation, close to any motor, the probability that the molarity has values $c_{\mathtt{oi}}\le c(t)<+\infty$ (with the initial condition $c(t^\prime)=c_{\mathtt{io}}$) is
\begin{equation}
S_\mathtt{o}(\tau)=\int_{c_{\mathtt{oi}}}^{+\infty} \breve{p}(c,t|c_{\mathtt{io}},t^\prime)\,\upd c\;.
\label{eqn:TSPF}
\end{equation}
Both SPF, given by equations~\eqref{eqn:RSPF} and \eqref{eqn:TSPF}, must satisfy initial conditions $S_\mathtt{i}(0)=S_\mathtt{o}(0)=1$ and asymptotic conditions $S_\mathtt{i}(+\infty)=S_\mathtt{o}(+\infty)=0$. For a system of $n$ equivalent and autonomous flagellar motors, the probability that a bacterium runs or tumbles is
\begin{eqnarray}
F_\run &=& S_\mathtt{i}^{\,n}\;,\nonumber\\
F_\tumble &=& (S_\mathtt{i}+S_\mathtt{o})^n- S_\mathtt{i}^{\,n}\;,\label{eqn:SX}
\end{eqnarray}
respectively. The runtimes and tumble-times distributions are
\begin{equation}
f_\xmotion(\tau)=-\frac{\upd F_\xmotion}{\upd\tau\hspace{1ex}}\;,\label{eqn:fX}
\end{equation}  
where $\xmotion=\run\;\mathrm{or}\;\tumble$, which can be calculated in different ways. The time distributions with $n=1$ [equations~\eqref{eqn:SX} and \eqref{eqn:fX}] are known as first passage time densities (FPTD). This is the survival probability density where the concentration reaches a threshold value $c_{\texttt{io}}$ ($c_{\texttt{io}}$) at time $t$, with initial condition $c_{\texttt{oi}}$ ($c_{\texttt{oi}}$), corresponding to the CCW (CW) rotation mode. In the case $n = 1$, it is convenient to refer to equation~\eqref{eqn:fX} as rotation-time distributions of the modes CW and CCW, since this case does not apply to the multi-flagellated \textit{E. coli}, but does apply to each of its rotary motors. Figure~\ref{fig:chey-p} shows the numerical results of the CheY-P molarity (around any flagellar motor) as a function of time, obtained by integrating one of the Langevin equations \eqref{eqn:c(t)} with parameters $\beta =1$ and $\mu=\SI{2}{\micro\molar}$. Alternatively, it is possible to give an analytical approach near the steady state. Using the ansatz of variables separation, the solution of the Fokker-Planck equation \eqref{eqn:FPc} near the steady probability density is
\begin{equation}
p(c,t|c^\prime,t^\prime)=p_\mathrm{st}(c)+\sum_{j=1}^{+\infty} \,T_j(\tau)\,C_j(c)\;,
\label{eqn:TCc}
\end{equation} 
where it is straightforward to find that $T_j(\tau)\sim\ex^{- \alpha_{j}\tau}$, with $\alpha_{j}>0$. The SPFs [from equations~\eqref{eqn:RSPF} and \eqref{eqn:TSPF}] are
\begin{equation}
S_\mathtt{y}(\tau)=\sum_{j=1}^{+\infty}s_{\mathtt{y}j}\,\ex^{- \alpha_{j}\tau}\;\label{eqn:Sx}
\end{equation}
(with $\mathtt{y}=\mathtt{i,o}$), where the coefficients are $s_{\mathtt{i} j}=\int_0^{c_{\mathtt{io}}}C_j(c)\,\upd c\;$ for the run and $s_{\mathtt{o}j}=\int_{c_{\mathtt{oi}}}^{+\infty} C_j(c)\,\upd c\;$ for the tumble. Note from the equation~\eqref{eqn:Sx} that the initial condition $S_\mathtt{y}(0)=1$ allows to establish the sum rule for the coefficients $\sum_{j=1}^{+\infty}s_{\mathtt{y}j}=1$ and that the asymptotic condition $S_\mathtt{y}(+\infty)=0$ is verified. It is easy to see through equations~\eqref{eqn:SX} and \eqref{eqn:Sx} that the run- and tumble-time distributions (given by equations~\eqref{eqn:fX}) near the steady state are linear combinations of decreasing exponential functions, which is a conclusion that is consistent with the experimental observations \cite{Berg1972,Korobkova2004}. This conclusion includes the rotation-time distributions of the CW and CCW modes. However, the variable separation approach is inadequate to establish the early time distributions. Instead, by integrating Langevin's equations~\eqref{eqn:c(t)}, we show that time distributions are increasing functions from zero, reaching an early maximum and then decaying as mentioned above, as we show in Figure~\ref{fig:TimeDistr}. The first experimentally measurements (\textit{circa} 1972) on run- and tumble-time distributions for \textit{E. coli} showed exponential decays although they did not show either maxima at early times or zero value at initial time \cite{Berg2004}. This is reasonable because both features happen at very short times $\lessapprox\SI{0.04}{\second}$. The maxima of the time distributions at early times for \textit{E. coli} have been first experimentally observed fifteen years ago \cite{Korobkova2004}. The behaviour described here has also been observed in time distributions of runs and reverse runs performed by \textit{V. alginolyticus} with a single flagellum \cite{Xie2011}.
\begin{figure}[ht!]
\centering
\includegraphics[scale=0.25]{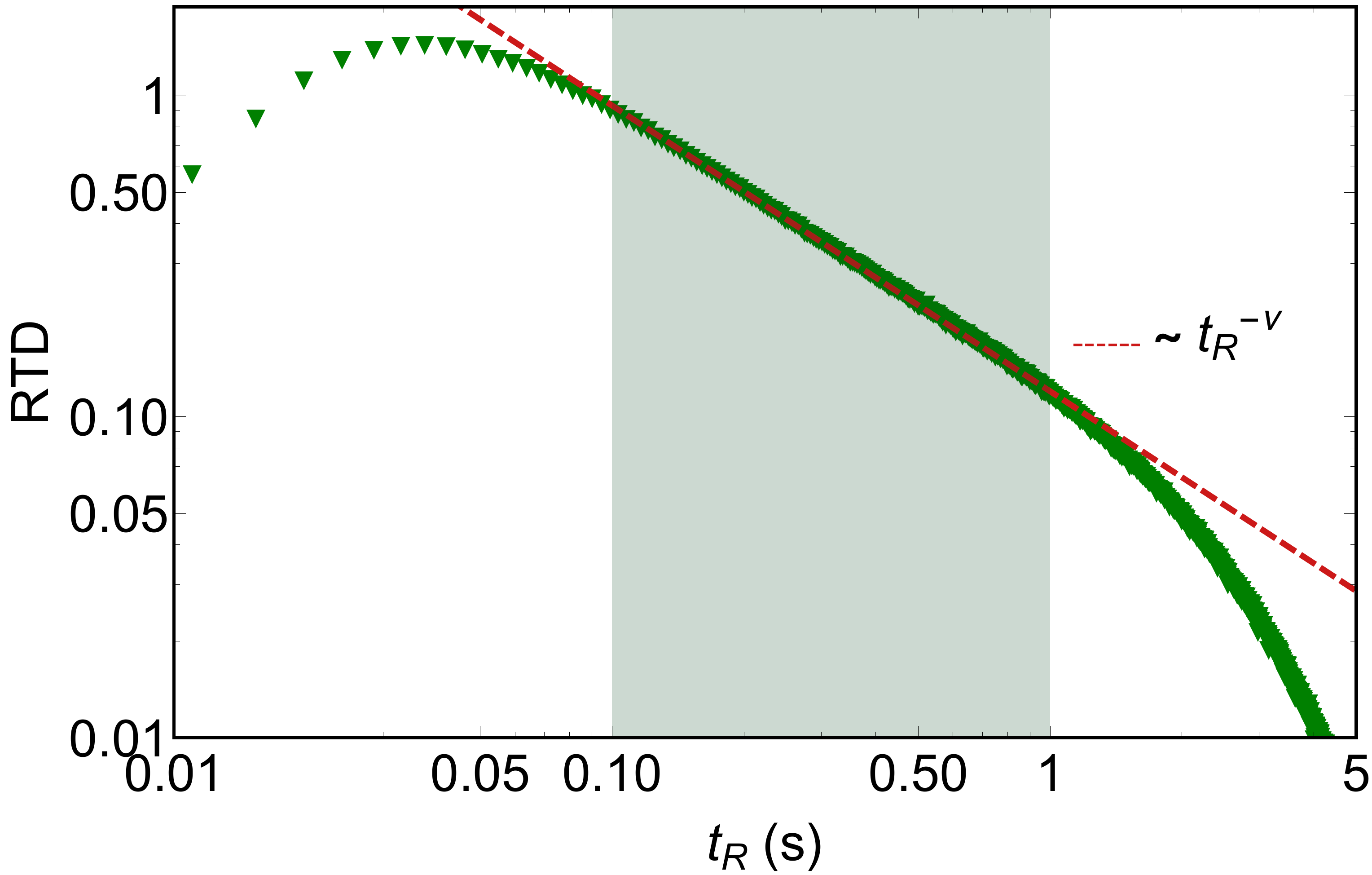}
\caption{The log-log plot shows the runtime distribution (RTD) as a function of runtime $t_\run$ (with green triangles), using the same data as in the right plot of Figure~\ref{fig:TimeDistr}. In the interval $[0.1,1]$ we observe a power-law behaviour, \textit{i.e.} $f_\run(\tau)\sim\tau^{-\nu}$ with exponent $\nu=0.89$ (shown by the red dashed line). For $\tau\gg 1$, we observe an exponential cutoff.} 
\label{fig:loglog-runtime-dist}
\end{figure}

Based on the experimental observation and our theoretical conclusions, the time distributions of runs and tumbles can be described by
\begin{equation}
f_\xmotion(\tau)=\sum_{k=1}^{+\infty}\,B_{\xmotion_k}\,\ex^{-\beta_{\xmotion_k}\tau}\;.
\label{eqn:fit-time-distr}
\end{equation}
Both plots of Figure~\ref{fig:TimeDistr} show numerical data fits by using equation~\eqref{eqn:fit-time-distr}, which is a good approximation even at early times. The left plot shows that, after the maximum, the tumble-time distribution can be fitted by a single exponential function. In turn, Figure~\ref{fig:loglog-runtime-dist} shows that the numerical data of the runtime distribution can be fitted into at least one decade with a power-law function. After this behaviour, an exponential cutoff is observed. The runtime distribution
\begin{eqnarray}
f_\run(\tau)\sim\biggl\{
\begin{array}{ll}
\tau^{-\nu}	&\hspace{2ex} \tau\in [a,b]\\
\ex^{-\lambda\tau}	&\hspace{2ex} \tau\gg b\;,\label{eqn:power-law-distr}
\end{array}
\end{eqnarray}
where $\nu$ is the power-law coefficient, $\lambda$ is the cutoff coefficient, and $[a,b]$ is the time interval where the power law is observed. The experimental observation and our model agree that the tumble-time distributions are mono-exponential. With the same model we show that multi-exponential runtime distributions fit the experimental measurements very well. On the other hand, the runtimes distributions have been adjusted, by us and other authors, with power laws for one or more decades \cite{Korobkova2004}. This mismatch should not be surprising, as it is possible to fit in a certain range power-law distributions with multi-exponential functions \cite{Bochud2007}. Conversely, a weighted sum of $N$-exponential functions may result in a power law at some interval, with particular weights $B_{\run_k}$ and  characteristic times $\beta^{-1}_{\run_k}$, with $k=1,\dots,N$. This suggests that the CheY-P molarity fluctuations during the runs, assuming that the internal signaling system is unstimulated, self-organize the flagellar system reaching runtime distributions with power-law behaviour at time range. The sleep- and wake-stage distributions show similar behaviours to those described for tumble and run respectively: the disruptive-sleep duration follows exponential distributions, while the wake duration is self-organized with power-laws distributions \cite{ChuShore2010}. These behaviours resemble the dynamics seen in some models of self-organized criticality (SOC): avalanche-time distributions follow power laws, while quiessence-time distributions can be exponential \cite{Buceta2011}. However, our system is out of criticality, since the mean runtime is finite. If, when changing the molarity parameters, the power-law range increases, then the mean runtime also increases. If the power-law range and mean runtime go to infinity, the system reaches a steady critical state \cite{Yang2004}, which does not happen in our system. From our simulations we have observed that the power-law behaviour of the runtime distribution is lost when the internal signaling system begins to be stimulated, preserving the multi-exponential behaviour.

The ultrasensitive property of flagella motors can be quantitatively described by a Hill equation \cite{Weiss1997} for the response $\theta$ as a function of the steady CheY-P molarity $\mu$. The response of an individual flagellar motor is defined as the time fractions in which the motor rotates in one direction and the other, referred to as CW bias or CCW bias and denoted here $\theta_{\CW}$ or $\theta_{\CCW}$, respectively. The Hill parameters are the steady molarity producing half response $\mu_{_\mathrm{A}}$ and the Hill coefficient $\Hill$, which measures the steepness of the response. With our model, varying the steady molarity $\mu$, we study the CW-bias change of each motor. As in this work we focused on explaining run-and-tumble time distributions for unstimulated scenarios, we do not take into account the adaptive methylation pathway. This is equivalent to use \mbox{CheR-CheB} mutants in ultra-sensitivity experiments. Thus, our changes in the steady molarity $\mu$ corresponds to changes in Chey-P induced by plasmids at different IPTG concentrations in bead assays \cite{Cluzel2000}. We average CW bias for each flagellar motor after several minutes, time scale much bigger than motor adaptation through turnover of C-ring proteins \cite{Yuan2012}. Under this assumption, the two threshold values represent the steady state number of FliM subunits of each motor. The left plot of Figure~\ref{fig:bias-vs-mu} shows how our model reproduces accurately the ultrasensitive property of a flagellar motor, with fit Hill-parameters of the experimental and simulation data very close to each other. 
\begin{figure}[ht!]
\centering
\includegraphics[scale=0.52]{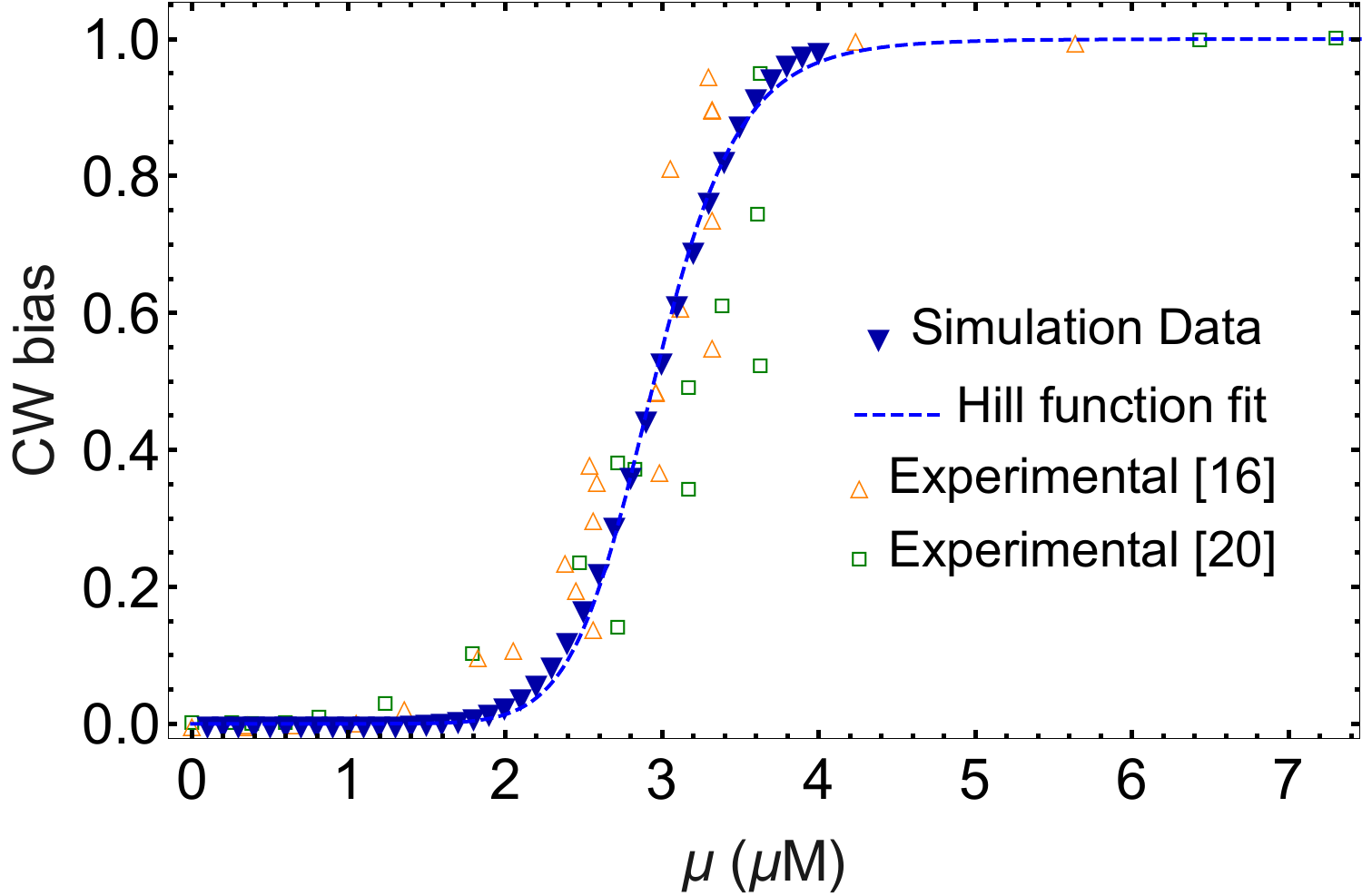}\hspace{2ex}
\includegraphics[scale=0.52]{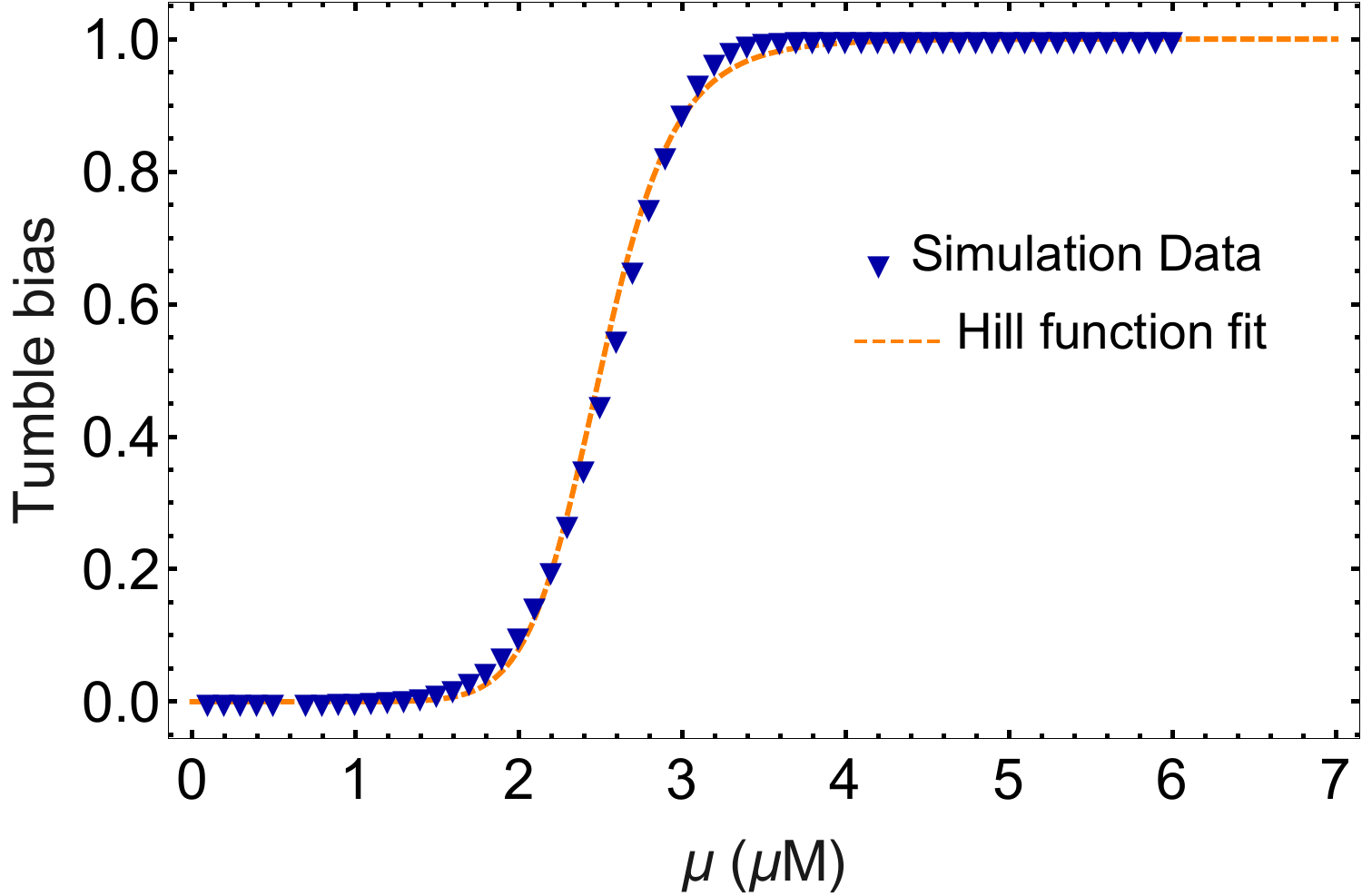}
\caption{Left: CW bias as a function of the equilibrium molarity $\mu$. Experimental and simulations data agree in the individual response of the flagellar motors \textit{vs} variations in CheY-P molarity. Using the Hill equation to fit our simulation data (with blue triangles down) we found that the molarity producing half CW-bias is $\mu_{_\mathrm{A}}=\SI{2.95}{\micro\molar}$ and the Hill coefficient is $\Hill=11.0$ (see blue dashed line). On the other hand, experimental data (with green squares) have been fitted with Hill parameters $\mu_{_\mathrm{A}}=\SI{3.1}{\micro\molar}$ and $\Hill=10.3\pm 1.1$ \cite{Cluzel2000}, which were confirmed with data from other measurements (with red triangles up) \cite{Sourjik2002b}. Right: Tumble bias as a function of the equilibrium molarity $\mu$. Plot shows the simulation data (with triangles) for a system with three flagella motors. The fit with the Hill equation has the same coefficient as that of the left plot although here $\mu_{_\mathrm{A}} = \SI{2.6}{\micro\molar}$\, (see blue dashed line).} 
\label{fig:bias-vs-mu}
\end{figure}

For a multi-flagellated bacterium, the response of the flagellar system is defined as the fraction of times in which the bacterium runs or tumbles, called run bias or tumble bias \cite{Mears2014} and denoted here $\theta_\run$ or $\theta_\tumble$, respectively. In addition, as we see in the right plot of Figure~\ref{fig:bias-vs-mu}, the tumble bias also shows ultrasensitive response under variations in CheY-P molarity, similar to CW bias, where its sigmoid curve is shifted left respect to the CW-bias one. Consequently, while $\mu_{_\mathrm{A}}$ is a function of the flagella number $n$, the Hill coefficient $\Hill$ is invariant. This implies that a higher number of flagella motors allows each bacteria to decrease the operational steady molarity leaving invariant the amplification of the flagella system. Using the alternative expression for the phenomenological Hill equation
\begin{equation}
\frac{1-\theta}{\theta}=\Bigl(\frac{\mu_{_\mathrm{A}}}{\mu}\Bigr)^{\!\Hill}\;,\nonumber
\end{equation}
it is easy to see that the right-hand side of this equation is equal to ${\langle t_\run\rangle}/{\langle t_\tumble\rangle}$ when $\theta=\theta_\tumble$ is the tumble bias or is equal to ${\langle t_{\CCW}\rangle}/{\langle t_{\CW}\rangle}$ when $\theta=\theta_{\CW}$ is the CW bias. Chemotactic response modifies cell behaviour changing the tumble frequency. Under the presence of attractants, such as amino acids like aspartate, \textit{E.coli} extends the runtimes although tumble-times remain almost constant \cite{Berg1972}. Because of this response, \textit{E. coli} has been described as an `optimist' \cite{Berg2004}, since when life gets better, it keeps swimming in the same direction. In the left plot of Figure~\ref{fig:mean-vs-bias} we show this optimist behaviour. For short tumble-bias $\theta_\tumble\lessapprox 0.15$, we show how the mean run- and tumble-times shift against constant chemotactic conditions. We found that runtimes increase with tumble bias following a power-law behaviour (see right plot of Figure~\ref{fig:mean-vs-bias}). In contrast, mean tumble-times remain almost constant around $\SI{0.1}{\second}$ as was observed \cite{Berg1972}. A similar behaviour is found for the mean CW- and CCW-times, which we do not show here because of its great similarity. For long tumble-bias $\theta_\tumble\gtrapprox 0.85$ we found the opposite effect, where mean tumble-time behaves like mean runtimes and vice versa. To the right of Figure~\ref{fig:mean-vs-bias}, the log-log plot shows that mean run- and tumble-times as functions of tumble- and run-bias, respectively. These mean times have a power-law behaviour for more than four decades. Surprisingly, the power-law exponent takes the same value regardless of the flagella number and matches the exponent value of the runtime distribution. 
\begin{figure}[ht!]
\centering
\includegraphics[scale=0.5]{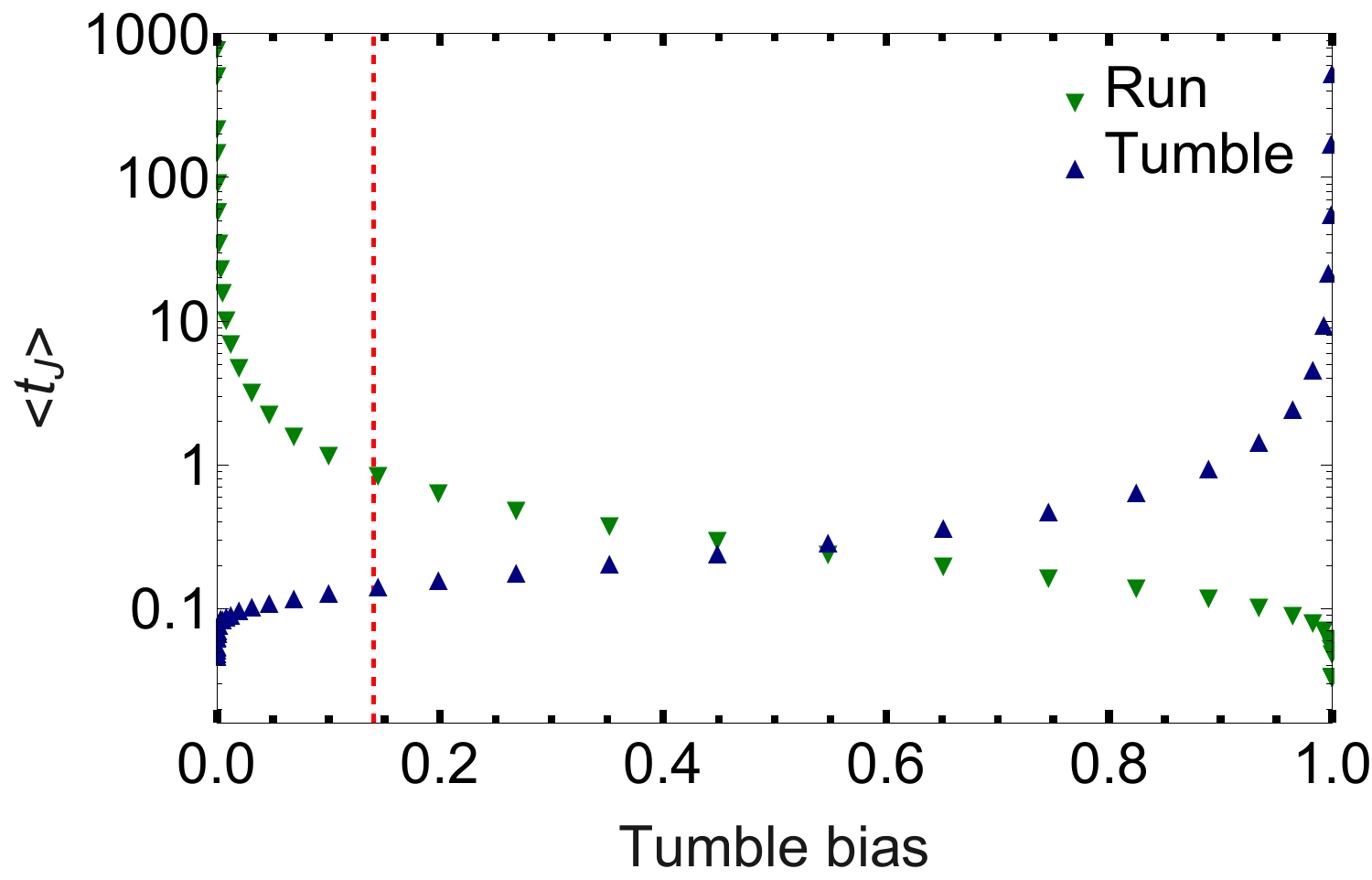}
\includegraphics[scale=0.49]{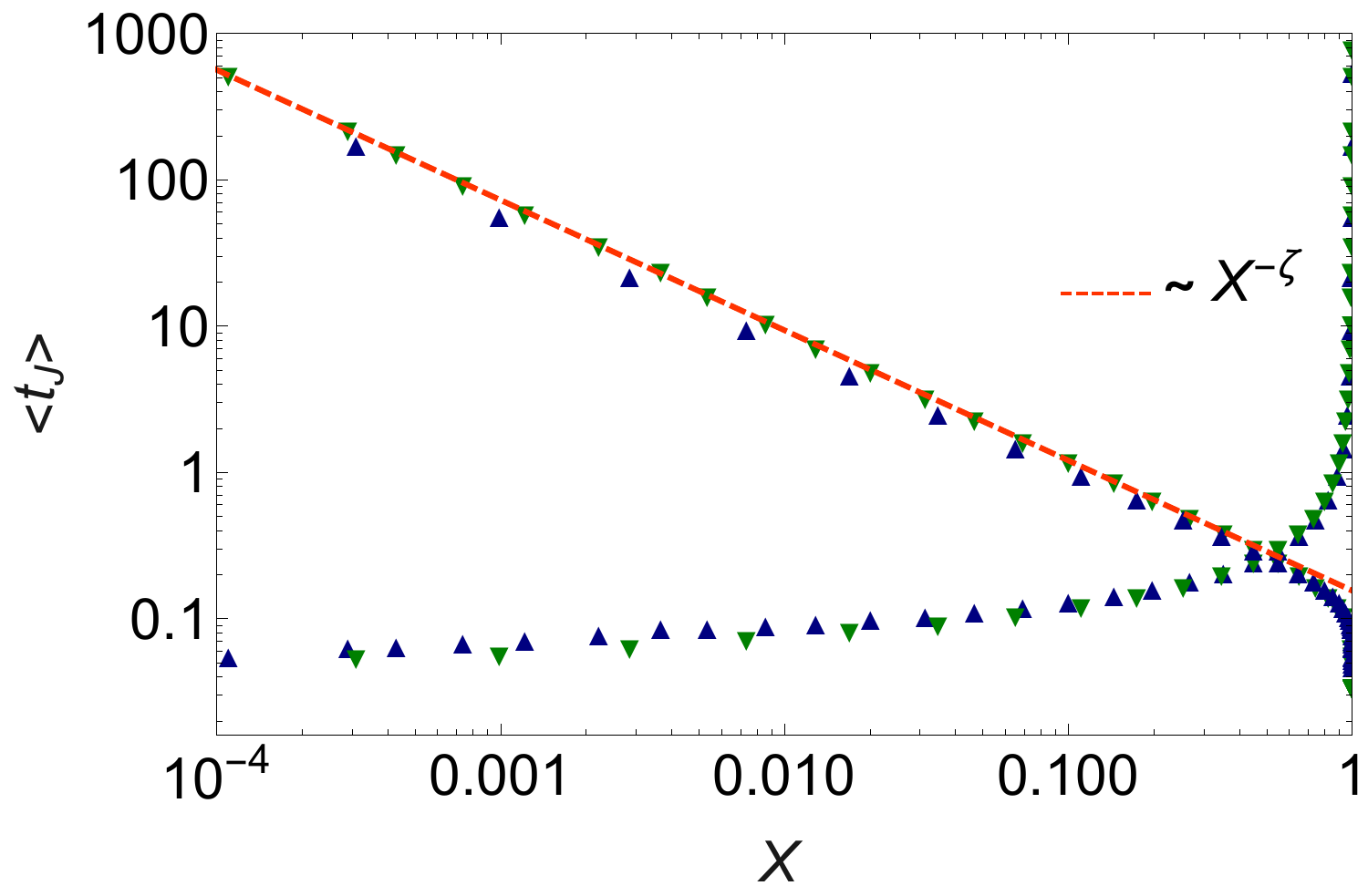}
\caption{Left: The semi-log plot shows the mean run- and tumble-times as functions of the  tumble bias, where $\mathtt{J}=\run,\tumble$. The vertical red dashed line shows tumble-bias value $\theta_\tumble\approx 0.15$ for unstimulated run-and-tumble. Low (high) tumble-bias results in long (short) runs and short (long) tumbles. Right: The log-log plots show the mean runtime (and mean tumble-time) as functions of the tumble bias $X=\theta_\tumble$ (and run bias $X=\theta_\run=1-\theta_\tumble$), plotted with green triangles down (and blue triangles up), respectively. The red dashed line shows the power-law behaviour with exponent $\zeta=0.89$.}
\label{fig:mean-vs-bias}
\end{figure}

Kinase activity have been measured via fluorescence resonance energy transfer (FRET) technique following the signal of \mbox{CheY-P-CheZ} complex \cite{Sourjik2002a,Tu2008,Shimizu2010}. From this measurements, phenomenological equations for kinase activity and methylation level has been proposed in references \cite{Tu2008,Shimizu2010}. Thus, we can link the steady CheY-P molarity $\mu$ with the kinase activity function $G$ through the linear relationship $\mu = k\,G(m,[L])$, where $k$ is a constant, [L] is the ligand concentration, and $m$ is the average methylation level of the receptors. We show that the two-threshold hypothesis is sufficient to reproduce the second step of amplification (see Figure~\ref{fig:bias-vs-mu}). The addition of the kinase activity allows the model to reproduce the first step of amplification. For \mbox{CheR-CheB} mutants, the methylation remains always at the same steady state; therefore, we can simplify the dependence of the kinase activity only to ligand concentration, \textit{i.e.} $G=G([L])$. We choose the steady state methylation so that the kinase activiy reaches the basal value for unstimulated conditions, which we maintain fixed. Through the inverse function $G^{-1}$  of the kinase activity, we can find the relationship $[L]=G^{-1}(\mu/k)$. This allows us to relate the tumble bias and CW bias with attractants concentration, as has been done experimentally \cite{Sourjik2002b}. Because one of our main goals is to reproduce the run-and tumble-distributions, we fixed the unstimulated tumble bias $\theta_\tumble=0.15$, which varies from strain to strain \cite{Berg1972}. Thus, we study the effect of attractans for tumble bias $\theta_\tumble\le 0.15$. This also impacts on the CW bias, which will always be smaller than tumble bias, being $\theta_{\CW}=0.04$ for unstimulated conditions. In left plot of Figure~\ref{fig:bias-vs-[L]} we show how the tumble bias is modified against flagella number. In this parametric plot we analyze the tumble bias as a function of the  CW bias under increasing $\mu$; that is, the enhancement in the response against flagella number. We found that increasing the number of flagella allows each bacteria to perform chemotaxis with lower concentrations of CheY-P. This same results have been found experimentally in reference \cite{Mears2014} for other \textit{E.coli} strain. Combining with results of Figure~\ref{fig:bias-vs-mu}, we conclude that more flagella allows lower CheY-P concentrations where the shape of the response will be invariant. Another way to study the impact of flagella number in chemoctactic response can be achieved by the linear relationship between $\mu$ and [L] through the inverse function $G^{-1}$ as we mentioned above. In right plot of Figure~\ref{fig:bias-vs-[L]} we found that a higher flagella number increases the bias, allowing higher amplification in the response, which is a consequence of the veto rule. Furthermore, more flagella increases the robustness of the response; \textit{e.g.} the CW bias of any flagellar motor drops to zero for $[L]\approx\SI{5}{\micro\molar}$ and the tumble bias of three-flagella system drops to zero for $[L]\gtrapprox\SI{10}{\micro\molar}$ (see right plot of Figure~\ref{fig:bias-vs-[L]}). The range of response for the ligand concentration agrees with the mean values used for chemotactic experiments \cite{Berg1972}. Thus, our simulations confirm the `gain paradox' \cite{Berg2004} with another possible amplification step and robustness enhancement of the chemotactic response induced by an increase of flagella number.

Finally, we remark that our conclusions do not depend of the flagella number used in the simulations. Recently, has been observed that flagella number thus not only vary from strain to strian but also is sensitive to growing media conditions \cite{Sim2017}. Thus, the parametrization used in our model depends of the strain and environment conditions used in reference \cite{Berg1972}; comparisons with other experimental setups must be done qualitatively and not quantitatively. The latter one requires reparametrization.

\begin{figure}[t!] 
\centering
\includegraphics[scale=0.335]{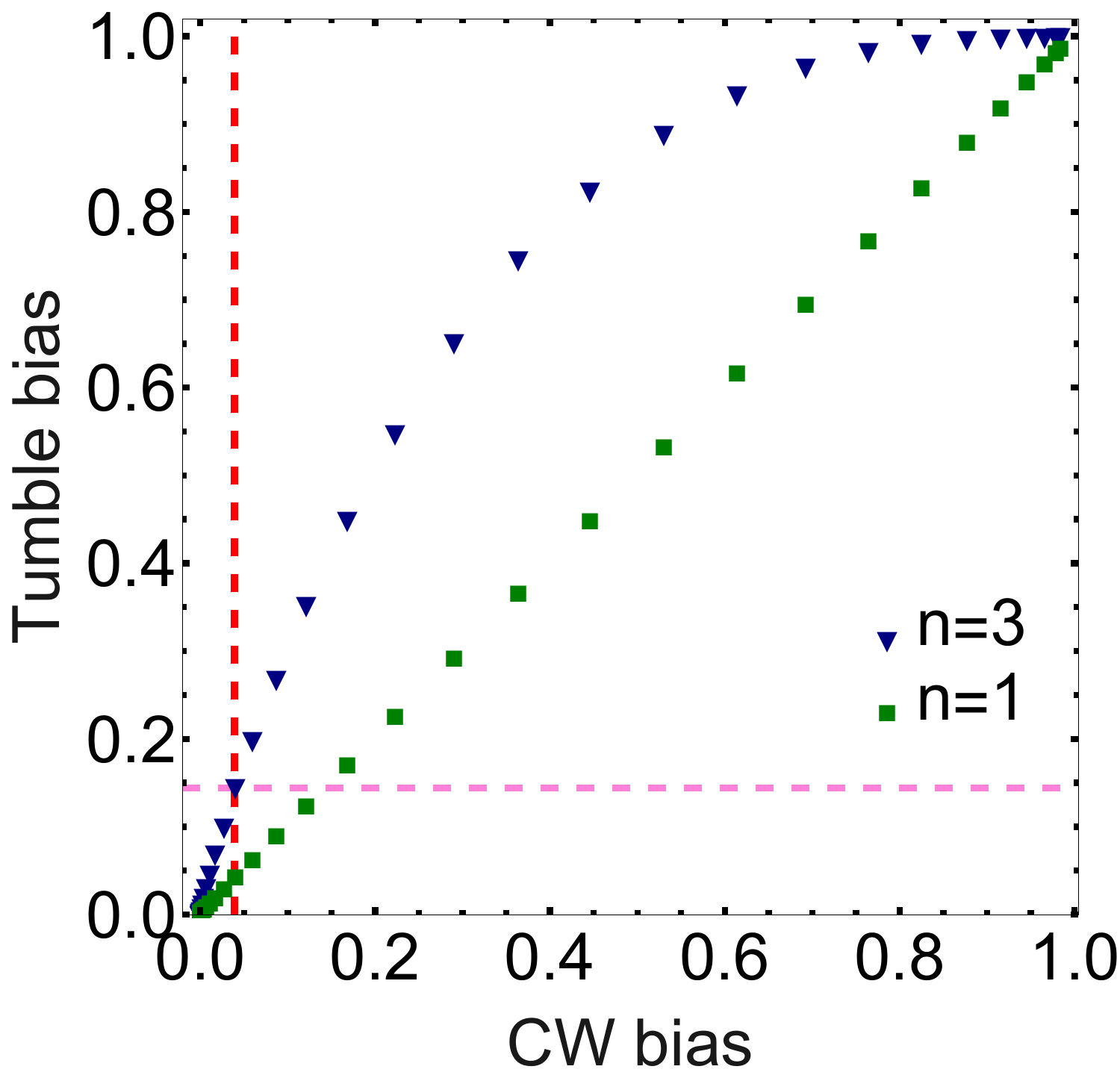}\hspace{4ex}
\includegraphics[scale=0.35]{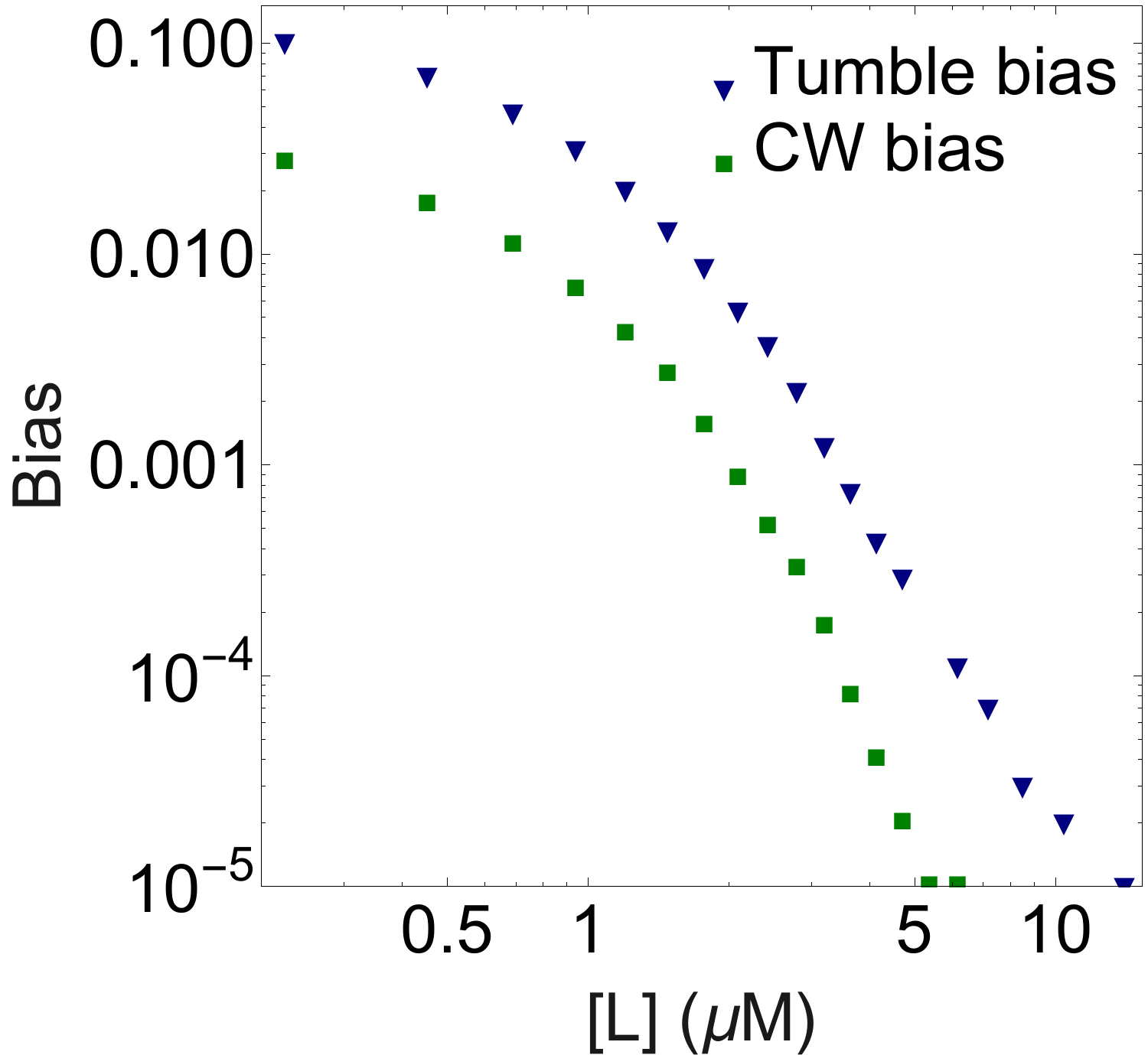}
\caption{Left: The plot of the tumble bias as a function of the CW bias shows the bias deviation with the increase of the flagella number. The intersection of the dashed lines shows the bias values $\theta_{\CW}=0.04$ and $\theta_\tumble=0.15$ for an unstimulated bacteria with three flagella. Right: Log-log plot of the tumble- and CW-bias as functions of the ligand concentration [L]. The tumble-bias data shown correspond to a three-flagella system of our simulation. When \mbox{$\mathrm{[L]}\rightarrow 0$} we recovered the motion biases from unstimulated bacteria. When the CW bias dropping to zero the tumble bias is about an order of magnitude greater, which shows the robustness of the bias response.}
\label{fig:bias-vs-[L]}
\end{figure}

\section{Conclusions\label{sec:conclusions}}

The last five decades of research on bacterial systems, whether as a single bacterium or a colony, account for the challenge of explaining phenomena that are dependent on different spatial and temporal scales. When we study the individual movement of a flagellated bacterium in a non-chemotactic media free from interactions with other bacteria or boundaries, we observe that the motion modes on a microscopic scale are determined by the rotation direction of the flagella that are hooked to molecular rotary motors of nanometric diameter. Each motor makes transitions in the rotational direction (CW or CCW) in a few hundredths of a second \cite{Berg2004}, while the time of each motion mode is at least an order of magnitude greater (\textit{e.g. E. coli} can have runtimes of the order of a second and tumble-times of the order of a tenth of a second), and a motion-mode sequence can last several tens of minutes. The statistical properties of microscopic movement does not reveal information about the internal processes of the bacteria if we do not link phenomena at different scales. The CW bias as a function of the equilibrium molar concentration of CheY-P (in the vicinity of the rotating molecular motor) can be established experimentally, although it has not been conveniently modeled until today. The run- and tumble-times distributions, and therefore the mean run- and tumble-times, observed microscopically, must be closely linked to the CheY-P molarity in the vicinity of each motor. This is one of the most relevant connections, although it is not the only one. This paper presents a simple model for \textit{E. coli} that tries to establish how phenomena that occur at a nanometric scale in the cytoplasm give rise to motile behaviours on a microscopic scale. 

In Section~\ref{sec:two} we present a system of Langevin equations for the molarity of CheY-P in the vicinity of each rotating molecular motor. We assume that when \textit{E. coli} runs or tumbles in an isotropic environment, its internal signaling system is unstimulated, and as a result, the \mbox{CheY-P} molar concentration is stable. We also assume that the transitions between CCW and CW rotations occur due to fluctuations in CheY-P concentrations in the presence of two threshold values, a hypotheses introduced about two decades ago \cite{Morton1998,Bren2001}. Inspired by the theory of first passage time density, which we show is valid for a uniflagellated bacterium, we propose a model for \textit{E. coli}, valid for other peritrichous enteric bacteria, \textit{e.g. Salmonella enterica}, assuming a veto hypothesis \cite{Vladimirov2010,Sneddon2012,Mears2014}. The most relevant conclusion of our internal signaling model is that, near the steady state, the run- and tumble-time distributions, as well as the rotation-time distributions of the CW and CCW modes, are linear combinations of decreasing exponential functions. These results are in full agreement with the pioneering experimental observations on time distributions made by Berg and Brown \cite{Berg1972}. The formalism presented cannot predict the maximum at short times, which has been recently observed experimentally, as well as it cannot predict the behaviour at early times, which we describe through the numerical integration of Langevin equations. This complex response of the flagellar system can be attributed, on the one hand, to the two regulatory thresholds of the CW-CCW transitions and, on the other hand, to the veto hypothesis. There has been a long debate about whether these runtime distributions are multi-exponential or power law after the maximum. Here we show that the tumble-time distribution is mono-exponential and the runtime distribution is multi-exponential, which can be precisely fitted with a power law in certain time range only if the internal signaling system is unstimulated. We conclude that, during runs without stimuli, the CheY-P molarity fluctuations close to rotatory motors self-organize the flagellar system out of criticality during some time range to reach runtimes that follow a power-law-like distribution. The runtime distribution in later times, after those of power-law behaviour, presents an exponential cutoff, which is a sign that the flagellar system loses its self-organization. Similar power-law behaviours with exponential cutoff have been established by Tu and Grinstein \cite{Tu2005} for CCW-rotation-time distributions from a linear mean-field model for the concentration of \mbox{CheY-P}. The mono-exponential behaviour of the tumble-times distribution and the power-law-like behaviour of the runtimes distribution are lost when the system is stimulated, preserving only the multi-exponential behaviour of both distributions. Nevertheless, the run- and tumble-time distributions of \textit{E. coli} swimming unstimulated gain importance in chemotactic scenarios. When the receptor kinase activity is modified by external chemical stimuli, a slow adaptation process begins through the response regulator CheB. After time scales in the range of minutes, the CheY-P molarity shifts back to the unstimulated stable value. In consequence, after some minutes, the system returns to the default run- and tumble-time distributions even under invariant chemotactic conditions. Our model also allows to accurately describe the ultrasensitive response of flagellar motors, as well as the entire flagellar system. The responses (CW bias and tumble biases) are sigmoid functions of steady molarity similar to each other, which shift to the left when the flagella number increases. A higher flagella number produces an increase in amplification and robustness of the chemotactic response, being another possible amplification step in the signaling system pathway. Thus, the two-threshold hypothesis \cite{Bren2001,Yuan2012} combined with the veto rule \cite{Mears2014} allow to explain a variety of phenomena for stimulated and unstimulated conditions described through the past decades. We finally conclude that the knowledge of the time distributions of each motion mode as a function of nanoscopic parameters is essential when studying other observables usually measured in the laboratory at a microscopic scale.

\section*{Acknowledgements}
This work was partially supported by Consejo Nacional de Investigaciones Cient{\'i}ficas y T{\'e}cnicas (CONICET), Argentina, PIP 2014/16 $\mathrm{N^o}\!$ 112-201301-00629. R.C.B. thanks to M. Semp\'e for her suggestions on the final manuscript.
\bibliography{FHB3-biblio}

\end{document}